\documentclass[fleqn,10pt]{SelfArx} 

\definecolor{color1}{RGB}{0,0,90} 
\definecolor{color2}{RGB}{0,20,20} 

\usepackage{hyperref} 
\hypersetup{hidelinks,colorlinks,breaklinks=true,urlcolor=color2,citecolor=color1,linkcolor=color1,bookmarksopen=false,pdftitle={Title},pdfauthor={Author},urlcolor=blue}

\usepackage{graphicx}
\usepackage[square]{natbib}
\usepackage{amsmath}
\usepackage{multirow}
\usepackage{subcaption}
\usepackage{pdflscape}
\usepackage[final]{pdfpages}

\usepackage{graphicx}
\usepackage{natbib}
\usepackage{amsmath}
\usepackage{amsfonts}
\usepackage{multirow}
\usepackage{tikz}
\usetikzlibrary{calc}

\usepackage{mathtools,amssymb,lipsum}

\usepackage{cuted}
\usepackage{boondox-calo}

\usepackage{calligra}

\newcommand{\mbf}[1]{\mathbf{#1}}

\DeclareMathAlphabet{\mathcalligra}{T1}{calligra}{m}{n} \DeclareFontShape{T1}{calligra}{m}{n}{<->s*[2.2]callig15}{}
\newcommand{\sr}{\ensuremath{\mathcalligra{r}}}

\newcommand{\bs}[1]{\boldsymbol{#1}}

\newcommand{\n}[1]{\mathrm{#1}}

\newcommand{\ud}{\mathrm{d}}

\JournalInfo{Published in IEEE Transactions on Magnetics, 62, 7000206, 2026} 
\Archive{\href{https://doi.org/10.1109/TMAG.2025.3632626}{DOI: 10.1109/TMAG.2025.3632626}} 

\PaperTitle{The magnetic scalar potential for a rectangular prism} 

\Authors{Berian James$^{1}$, Stefan Pollok$^{1}$, Jes Frellsen$^{2}$, Rasmus Bjørk$^{1,*}$} 
\affiliation{$^1$\textit{Department of Energy Conversion and Storage, Technical University of Denmark - DTU, DK-2800 Kgs. Lyngby, Denmark}} 
\affiliation{$^2$\textit{Department of Applied Mathematics and Computer Science, Technical University of Denmark - DTU, DK-2800 Kgs. Lyngby, Denmark}} 
\affiliation{*\textbf{Corresponding author}: rabj@dtu.dk} 

\Keywords{} 
\newcommand{\keywordname}{Keywords} 

\Abstract{We analytically solve Poisson's equation for the magnetic scalar potential generated by a uniformly magnetized rectangular prism and determine a closed-form solution for the magnetic scalar potential given only in terms of arctan and natural logarithmic functions. We show that the magnetic scalar potential can be written as a demagnetization vector, containing all the geometric information, multiplied with the magnetization, analogous to demagnetization tensors. We validate the derived analytical expression for the magnetic scalar potential by comparing with a finite element simulation and show that these agree perfectly. We finally extend the concept of the demagnetization vector and tensor, which contains the geometric information for the source generating the potential, to gravitational objects.}

\begin{document}

\flushbottom 

\maketitle 


\thispagestyle{empty} 

\section{Introduction}
Magnetic fields are a key part in numerous scientific and technological areas, from fusion reactors to electric motors in cars. The magnetic field from a dipole has been known since the days of Maxwell, but the magnetic field generated by more geometrically complicated objects can be difficult to determine analytically. Knowing these would be advantageous both experimentally and numerically, as the field can then be evaluated to arbitrarily high spatial precision, which is crucial in a range of scientific fields. For example in micromagnetism the magnetic field has to be evaluated on a highly resolved grid of small magnetic domains to allow for the assumption of uniform magnetization in each of these elements \cite{exl2020micromagnetism,bjork_explaining_2023}. In magnetohydrodynamics a high enough spatial resolution in the magnetic field is needed to predict charged fluid particle trajectories \cite{mackay2006divergence}, likewise for particle physics experiments \cite{bernauer2016measurement}. In robotics, navigation by magnetic field benefits vastly from high spatial resolution \cite{le20123, solin2018modeling} and finally this is also the case for nuclear magnetic resonance applications, where a highly homogeneous magnetic field is required around the sample \cite{moresi2003miniature,raich2004design}.

However, calculating the magnetic field can be both analytically and numerically hard. Instead, an approach is to consider the magnetic scalar potential, which is defined such that the gradient of this scalar potential is the magnetic field. The scalar potential can be defined if there are no free currents in a magnetic system. An additional benefit of knowing the magnetic scalar potential is that within physics-informed machine learning there are multiple advantages to utilizing scalar potentials to build physical properties directly into the networks \cite{pollok_magnetic_2023}. First of all, the field obtained from the gradient of a scalar potential is conservative. Secondly, for a neural network to learn the potential only a single loss function is needed \cite{solin2018modeling}, which means that balancing multiple losses, e.g. different field components or conservation properties, is not needed \cite{Bischof_2025}.

The magnetic scalar potential is known for a magnetic dipole and a uniformly magnetized sphere, but interestingly not for a uniformly magnetized rectangular prism. The magnetic field generated by a rectangular prism was first considered extensive in Ref. \cite{Rhodes_1954}, where the demagnetization interaction energy, which is the spatial integral of the magnetic scalar potential over a rectangular prism, is derived \cite{Hubert_book}. However, Ref. \cite{Rhodes_1954} only reports the final calculated interaction energy and not the scalar potential, which the authors must have considered as only an intermediate result. This is also the case in Ref. \cite{Schabes_2003} who also calculates an exact analytic formula for the magnetostatic interaction energy of a three-dimensional array of cubes. Again, the starting point is the magnetic scalar potential, and again this is never explicitly derived and only its volume average over another rectangular region is reported.

The magnetic scalar potential is also used to obtain the magnetic field, which is the gradient of the potential. For a homogeneously magnetized rectangular prism, the magnetic field was first obtained in Ref. \cite{joseph_1964} who calculated the magnetic field (expressed as the demagnetization tensor) exactly by taking the gradient of the magnetic scalar potential. However, in analytically calculating the field one can interchange the gradient operator and the integral operator as these are with respect to the coordinates of the evaluation point and of the magnetized source, respectively. This trick was utilized in Ref. \cite{joseph_1964}, a trend followed in later demagnetization tensor calculations for other geometries \cite{Smith_2010,Nielsen_2020,Slanovc_2022} and thus these calculations do not analytically calculate the magnetic scalar potential, even though they determine its gradient, the magnetic field. 

It is also worth noting that demagnetization factors for rectangular prisms have also been considered extensively in literature. In Ref. \cite{Joseph_1967} the fluxmetric (or ballistic) demagnetizing factor for a rectangular prism is calculated, which is defined as the ratio of the average demagnetizing field to the average magnetization at the midplane perpendicular to a chosen axis. In Ref.  \cite{Aharoni_1998} the magnetometric demagnetization factor, which is defined as the average demagnetizing field within the body to the average magnetization, is calculated, and in Ref. \cite{Fukushima_1998} the average field generated over another rectangular prism is calculated. All of these calculations however start with the already known magnetic field of a rectangular prism \cite{joseph_1964,Smith_2010} and thus do not derive the magnetic scalar potential either.

In this work, we explicitly calculate the magnetic scalar potential for a uniformly magnetized rectangular prism and compare this to a numerical model to validate the derived analytical expressions. It remains an interesting observation that this is not known even though the integral of the potential \cite{Rhodes_1954,Schabes_2003} as well as its gradient \cite{joseph_1964,Smith_2010} and the integral of this gradient \cite{Joseph_1967,Aharoni_1998,Fukushima_1998} are already known. We also show that the magnetic scalar potential can be written as the product of the demagnetization vector, containing the geometric information, and the magnetization, similar to the demagnetization tensor known for calculating the magnetic field.

\section{Theory}
If there are no free currents in a magnetic system, as is e.g. the case for permanent magnets, the magnetic field vector, $\mathbf{H}$, is irrotational. For a single magnetized body $\Omega$ in the whole space, the magnetic field can therefore be expressed as the gradient of a scalar field, termed the magnetic scalar potential, $\phi_M$, as:
\begin{equation}
\mathbf{H}  = -{\nabla}\phi_M\label{eq:PhiM}~.
\end{equation}

The relation between the magnetic field, $\mathbf{H}$, the magnetic flux density, $\mathbf{B}$, and the magnetization, $\mathbf{M}$, is defined as:
\begin{equation}
\mathbf{B} = \mu_0(\mathbf{H} +\mathbf{M}) ~,
\label{eq:Hdef}
\end{equation}
where $\mu{}_{0}$ is the vacuum permeability. 

Using this relation in the definition for the scalar potential, taking the divergence, and remembering Gauss's law for magnetism, ${\nabla}{}\cdot{}\mathbf{B}=0$, results in a Poisson-like equation:
\begin{equation}
-{\nabla}\cdot{}(\mu{}_{0}\nabla\phi_M)=-\nabla\cdot{}(\mu{}_{0}\mathbf{M} )~.\label{Eq.Poisson01}
\end{equation}

We remark that $\mathbf{M}$ is assumed to be nonzero only within the region $\Omega$.
Thus, the latter equation must satisfy the following condition on the boundary of $\Omega$, termed the surface $S'$:
\begin{equation}
\mathbf{\hat{n}}\cdot[-\nabla\phi_M]_{S'}=\mathbf{\hat{n}}\cdot \mathbf{M}~, \label{Eq.Poisson02}
\end{equation}
where we have denoted with $\mathbf{\hat{n}}$ the outward normal on the surface $S'$ and with $[-\nabla\phi_M]_{S'}$ the jump of the vector field $\mathbf{H}=-\nabla\phi_M$ across $S'$. The above equations have the same form as those for the electrostatic field produced by assigned  (volume and surface) charge densities and lead to identify the volume and surface \emph{equivalent magnetic charge} densities $\rho_M=-\nabla\cdot\mathbf{M}$ and $\sigma_M=\mathbf{\hat{n}}\cdot \mathbf{M}$, respectively \cite{Rhodes_1954}.

The solution of Eqs. \eqref{Eq.Poisson01}-\eqref{Eq.Poisson02} for a homogeneously magnetized body, i.e. such that $\mathbf{M}$ is constant in $\Omega$ (and, consequently, $\rho_M=0$), is the scalar potential produced only by the surface charge density $\sigma_M=\mathbf{\hat{n}}\cdot \mathbf{M}$ and is given by \cite{Jackson}:
\begin{equation}
    \phi_M(\bs{r})=\frac{1}{4\pi} \oint_{S'}\frac{\mathbf{\hat{n}}(\mathbf{r}')\cdot\mathbf{M}(\mathbf{r}')}{\|\mathbf{r}-\mathbf{r}'\|}\ud S'.\label{eq:PhiMsolFlat}
\end{equation}
The above equation has two sets of coordinates. The coordinates marked with a $'$ are the coordinates of the face that creates the magnetic field, whereas the non-marked coordinates are to the point at which the field is evaluated. Once the magnetic scalar potential has been determined, the magnetic field can be determined from Eq. \eqref{eq:PhiM}.

\section{The magnetic scalar potential of a rectangular prism}\label{app:face}
We consider a uniformly magnetized rectangular prism with side lengths [$2a$, $2b$, $2c$], as is tradition \cite{joseph_1964,Smith_2010}, and which is centered at the origin. To derive the magnetic scalar potential, we first consider the magnetic scalar potential at a point $\mathbf{r} = (x, y, z)$ due to an element on the $x'$-face of the rectangular prism. This is given by the following integral across the differential area element on the face $dA' = dy'dz'$. On an $x'$-face the dot product of the normal vector and the magnetization is $\mathbf{\hat{n}}(\mathbf{r}')\cdot\mathbf{M}(\mathbf{r}')=M_x$. The distance vector is simply:
\begin{equation}
    \sr{}=\|\mathbf{r}-\mathbf{r}'\|=\sqrt{(x-x')^2+(y-y')^2+(z-z')^2}.\label{Eq.rdef}
\end{equation}

Setting $x'=a$ the magnetic scalar potential from the face according to Eq. \eqref{eq:PhiMsolFlat} is then:
\clearpage
\begin{strip}
\begin{eqnarray}
\begin{aligned}
\phi_{M,x'}(\mathbf{r}) = &\frac{1}{4\pi}M_x\int_{-b}^{b} \int_{-c}^{c} \frac{dy'dz'}{\sqrt{(x - a)^2 + (y-y')^2 + (z-z')^2}}\\
=\ &\frac{1}{4\pi}M_x \int_{-c}^{c} -\ln\left(y-y'+\sqrt{(x - a)^2 + (y-y')^2 + (z-z')^2}\right) \Bigr|^{y'=b}_{y'=-b}  dz' \\
=\ &\frac{1}{4\pi}M_x \Biggl[\Biggl( z' + (x-a)\arctan\left(\frac{z-z'}{x-a}\right) - (x-a)\arctan\left(\frac{(y-y')(z-z')}{(x-a)\sr{}}\right) \\ &+ (y-y')\ln\left((z-z') + \sr{}\right) + (z-z') \ln\left((y-y'\right) + \sr{}) \Biggr)\Biggr|^{y'=b}_{y'=-b}~\Biggr]\Biggr|^{z'=c}_{z'=-c} \\
=\ &\frac{1}{4\pi}M_x \Biggl[\Biggl( - (x-a)\arctan\left(\frac{(y-y')(z-z')}{(x-a)\sr{}}\right) \\ &+ (y-y')\ln\left((z-z') + \sr{}\right) + (z-z') \ln\left((y-y'\right) + \sr{}) \Biggr)\Biggr|^{y'=b}_{y'=-b}~\Biggr]\Biggr|^{z'=c}_{z'=-c}~, \label{Eq.PhimFull}
\end{aligned}
\end{eqnarray}
\end{strip}

where the last step follows from the fact that the first two terms in the expression, $ z' + (x-a)\arctan\left(\frac{z-z'}{x-a}\right)$ do not depend on $y'$ and so when the limits of integration for $y'=\pm b$ are evaluated, these will be removed as constants of integration. This leaves a function symmetric in the two coordinates we have integrated over, as it should be. In the above integral there is no constant of integration corresponding to a gauge of zero, equivalent to $\phi_{M}\xrightarrow[\sr{} \to \infty]{} 0$.

The integral expressions for the $y'$- and $z'$-faces are obtained by permuting the coordinates. This can be structured in terms of a function $F$, which encapsulates these contributions in a form that can be directly applied to calculate the potential, with $i$, $j$, and $k$ being equal to $x-x'$, $y-y'$, and $z-z'$ in a specified order. The function is: 
$$
F(i,j,k) = -i\arctan\left(\frac{jk}{i\sr{}}\right) + j\ln(k+\sr{}) + k\ln(j+\sr{})~,
$$
where, as above in Eq. \eqref{Eq.rdef}, the function $\sr{}(i,j,k)$ is given by: 
$$
\sr{}(i,j,k) = \|\mathbf{r}-\mathbf{r}'\| = \sqrt{i^2+j^2+k^2}~.
$$
With this function defined, the potential generated by a single $x'$-face located at $x'=a$ is given by 
\begin{eqnarray}
\phi_{x\text{-face}} &=& \frac{1}{4\pi}M_x F(x-a, y\mp b, z\mp c) \\
&=& \frac{1}{4\pi}M_x\bigl([F(x-a,y-b,z-c) - \nonumber \\ 
&& \qquad \quad \ \, F(x-a,y-b,z+c)] - \nonumber \\
&& \qquad \quad \; [F(x-a,y+b,z-c) - \nonumber \\
&& \qquad \quad \ \, F(x-a,y+b,z+c)]\bigr)~.
\end{eqnarray}

The contribution from both $x'$-faces is given by:
$$
\phi_{x\text{-faces}} = \frac{1}{4\pi}M_x \cdot [F(x-a, y\mp b, z\mp c) - F(x+a, y\mp b, z\mp c)]~,
$$
as the surface normal is negative for the $x'$-face located at $x'=-a$, corresponding to $F(x+a, y\mp b, z\mp c)$.

The contributions from the $y'$- and $z'$-faces can be calculated by permuting the coordinates in the argument to the $F$ function, reflecting the cube's symmetry and the orientation of the magnetization vector. The result is that for the $y'$-faces:
\begin{equation}
\phi_{y'\text{-faces}} = \frac{1}{4\pi}M_y \cdot [F(y-b, z\mp c, x\mp a) - F(y+b, z\mp c, x\mp a)]~,
\end{equation}
and for the $z'$-faces:
\begin{equation}
\phi_{z'\text{-faces}} = \frac{1}{4\pi}M_z \cdot [F(z-c, x\mp a, y\mp b) - F(z+c, x\mp a, y\mp b)]~.
\end{equation}

It follows that the complete magnetic scalar potential of a prism is given as:
\begin{equation}
\phi_{M}(\mathbf{r}) = \phi_{x'\text{-faces}} + \phi_{y'\text{-faces}} + \phi_{z'\text{-faces}}~. \label{Eq.potential_faces}
\end{equation}
We note that this expression is valid both inside and outside the prism. The derived analytical expressions are available through the MagTense framework \cite{MagTense}, available at \\ \href{https://www.magtense.org}{www.magtense.org}.

\subsection{Tensor notation and other geometries}
It is possible to formulate the magnetic scalar potential in vector notation as \cite{Boulanger_2024}:
\begin{equation}
\phi_{M}(\mathbf{r})=\mathbb{N}_\phi(\mathbf{r})\cdot{}\mathbf{M} ~, \label{Eq.final}
\end{equation}
where we introduce $\mathbb{N}_\phi(\mathbf{r})$ as the demagnetization vector for the magnetic scalar potential. This contains all the geometric information necessary for calculating the scalar potential.

Examining Eq. \eqref{Eq.potential_faces} it can be seen that we can write the demagnetization vector for the magnetic scalar potential as:
\begin{eqnarray*}
\begin{aligned}
\mathbb{N}_\phi = & \frac{1}{4\pi}
   \begin{bmatrix}
    F(x-a, y\mp b, z\mp c) - F(x+a, y\mp b, z\mp c) \\
    F(y-b, z\mp c, x\mp a) - F(y+b, z\mp c, x\mp a) \\
    F(z-c, x\mp a, y\mp b) - F(z+c, x\mp a, y\mp b) \\
    \end{bmatrix}~.
\end{aligned}
\end{eqnarray*}

The demagnetization vector plays an identical role to the demagnetization tensor, which is a familiar topic from magnetostatics and is defined for the magnetic field as $
H(\mathbf{r})=-\mathbb{N}(\mathbf{r})\cdot{}\mathbf{M}$. This $3\times{}3$ tensor is known analytically for several geometrical objects or tiles, such as a prism \cite{Smith_2010}, cylinder \cite{Nielsen_2020,Slanovc_2022}, elliptic cylinder \cite{Beleggia_2005}, tetrahedron \cite{Nielsen_2019}, ellipsoid \cite{osborn_1945,Tejedor_1995}, cylindrical shells \cite{Beleggia_2009} and hollow sphere \cite{PratCamps_2016} as well as several more. The average demagnetizing factor is also known for a range of geometries such as prisms and ellipsoids \cite{Joseph_1967,Aharoni_1998,Chen_2002,Chen_2005}.

The expression for the demagnetization tensor for the prism is the gradient of the scalar potential demagnetization vector derived in this work. We have omitted the minus sign in the definition of the demagnetization vector in Eq. \eqref{Eq.final}, as the magnetic field is defined as the negative gradient of the magnetic scalar potential.

We also note that it is possible to write down the demagnetization vector for the magnetic scalar potential for a range of other geometries for which the scalar potential is known. For a dipole, the magnetic scalar potential is given by: 
$$
\phi_{M,\n{dipole}}=\frac{1}{4\pi}\frac{\mathbf{r}\cdot{}\mathbf{m}}{r^3}~,
$$
where $\mathbf{m}$ is the magnetic moment. So the demagnetization vector is simply:
\begin{equation}
\mathbb{N}_{\phi,\n{dipole}}(\mathbf{r})=\frac{1}{4\pi}\frac{\mathbf{r}}{r^3}~.
\end{equation}

For a sphere, the magnetic scalar potential is given by:
$$
\phi_{M,\n{sphere}}=\frac{1}{3}\left(\frac{2R}{r+R+| r-R|}\right)^3\mathbf{r}\cdot{}\mathbf{m}~,
$$
so the demagnetization vector is:
\begin{equation}
\mathbb{N}_{\phi,\n{sphere}}(\mathbf{r}) = \frac{1}{3}\left(\frac{2R}{r+R+| r-R|}\right)^3\mathbf{r}~.
\end{equation} 
This expression is valid both inside and outside the sphere, as is also the case for the rectangular prism.

\subsection{Singular values}
The expression for the magnetic scalar potential for a face of a rectangular prism, Eq. \eqref{Eq.PhimFull}, has a number of singularities that must be considered. We consider a prism face at $x'=+a$. The singularities are described in the following. \\

\subsubsection{First term in Eq. \eqref{Eq.PhimFull}}
For $x=a$ the first term in the expression for the scalar potential, Eq. \eqref{Eq.PhimFull}, cannot be evaluated. However, taking the limit of $x \to a$ for the first term, we get:
$$
- (x-a)\arctan\left(\frac{(y-y')(z-z')}{(x-a)\sr{}}\right)\xrightarrow[x \to a]{} 0~,
$$
as $ \lim_{x \to \infty} \textrm{arctan}(x) = \pi/2$.

There is a special case when $y=y'$ or $z=z'$, i.e. when we consider the boundaries of the face. Besides the corner points, this limit is easily resolved as e.g. $(x-a)\rightarrow{}0$ as fast as $y-y' \rightarrow{}0$, resulting in a finite value of the arctan factor, and as this is multiplied with $(x-a)$, again the first term goes to zero.

The case of $x-a=0$, $y-y'=0$, $z-z'=0$, i.e.  $\sr{}\rightarrow{}0$, must also be considered. However, again we end up with the $\textrm{arctan}$ factor approaching a constant value, leaving the first term going to zero because of the $x-a$ prefactor. \\

\subsubsection{Second and third terms in Eq. \eqref{Eq.PhimFull}}
The second and third terms in Eq. \eqref{Eq.PhimFull} must also be considered, as these can also become singular. 

For $x=a$, $y=y'$ and $z=z'$, i.e. $\sr{}\rightarrow{}0$ the limit is: 
$$
(y-y')\ln\left((z-z') + \sr{}\right)\xrightarrow[\sr{} \to 0]{} 0~,
$$
as $ \lim_{x \to 0} x\textrm{ln}(x) = 0$.
The same applies to the third term.

Using the $ijk$ nomenclature introduced earlier, the limits for the singular values are given in Table \ref{Table.limits}, where it is seen that for all singular values, the considered term in Eq. \eqref{Eq.PhimFull} simply goes to zero.

\begin{table}[!t]
    \centering
    \caption{The limiting values for the magnetic scalar potential.}
    \begin{tabular}{cccc}
       & $i\textrm{arctan}\left(\frac{jk}{i\sr}\right)$  & $j\ln(k+\sr)$  & $k\ln(j+\sr)$ \\ \hline
      $i\rightarrow{}0$ & 0 & - & -\\
      $i\rightarrow{}0$, $j\rightarrow{}0$ & 0 & 0 & - \\
      $i\rightarrow{}0$, $k\rightarrow{}0$ & 0 & - & 0 \\
      $\sr\rightarrow{}0$  & 0 & 0 & 0\\
    \end{tabular}
    \label{Table.limits}
\end{table}

\section{Validation}
\begin{figure}[ht]
    \centering
    \includegraphics[width=\linewidth]{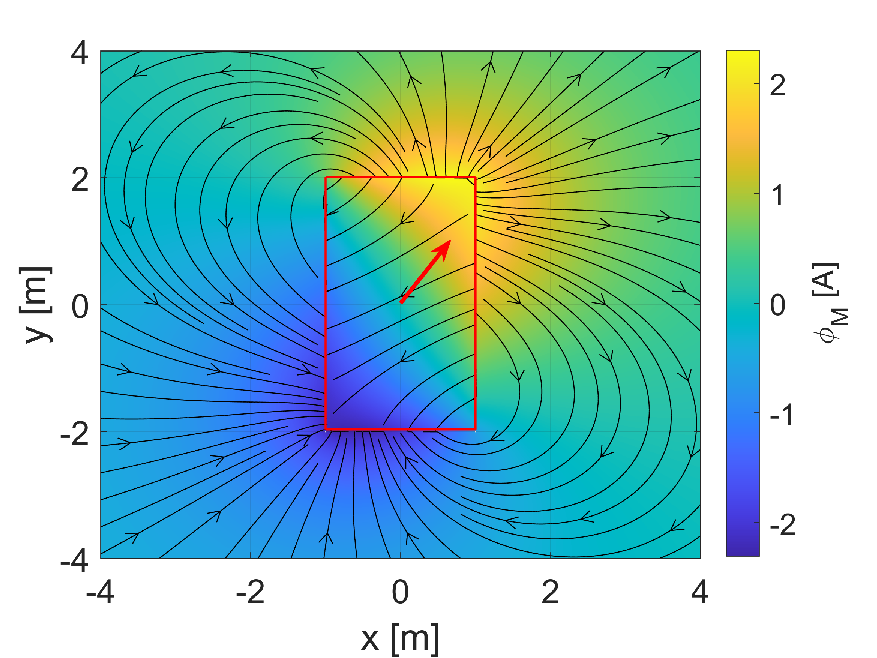}
    \caption{The magnetic scalar potential, $\phi_M$, and the magnetic field lines generated by a prism with side lengths $[2,\,4,\,6]$ m and magnetization of $M = [2,\,  3,\, -4]$ A/m in the $xy$-plane at $z=0$ m. The magnetization direction is indicated by the red arrow in the center of the prism, and the sides of the prism are shown with red lines.}
    \label{fig:prism-slice}
\end{figure}

We consider a prism with side lengths $[2,\,4,\,6]$ m, i.e. $a = 1$ m, $b = 2$ m, $c = 3$ m, centered at the origin. We consider a magnetization of $M = [2,\,  3,\, -4]$ A/m. In Fig. \ref{fig:prism-slice}, we show the magnetic scalar potential and the magnetic field lines computed from the prism in a slice in the $xy$-plane at $z=0$ m.

\begin{figure}[ht]
    \centering
    \includegraphics[width=\linewidth]{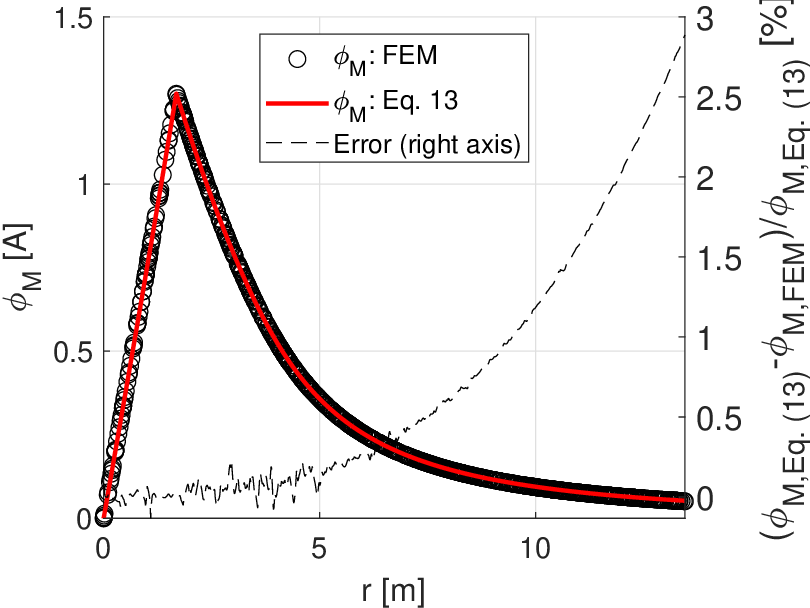}
    \caption{The magnetic scalar potential, $\phi_M$, generated by a prism with side lengths $[2,\,4,\,6]$ m and magnetization of $M = [2,\,  3,\, -4]$ A/m along the line $[0,\, 0,\, 0]\rightarrow[8,\, -6,\, -9]$ m as given by Eq. \eqref{Eq.final} and as computed using finite element modeling (FEM). The difference between these, in percent, is shown on the right axis.}
    \label{fig:prism-line}
\end{figure}

To validate the analytical expression for the magnetic scalar potential, we compare the scalar potential with that calculated using the finite element framework COMSOL \cite{comsol}. We evaluate the scalar potential from the origin to a point located at $P = [x,\, y,\, z] = [8,\, -6,\, -9]$ m. In Fig. \ref{fig:prism-line}, we show the magnetic scalar potential evaluated on a line ranging from the origin to the point $P$ specified above as computed using COMSOL and as computed using Eq. \eqref{Eq.final}, as well as the difference between these. As can be seen, there is a perfect agreement between the analytical expression and the finite element calculations, and the small error can be attributed to the numerical precision of the finite element model.

\section{Discussion}
As stated previously, in magnetostatics, the demagnetization tensor for the magnetic field is well known and has been computed for a number of geometries.

However, outside magnetism, the gravitational potential can also be defined as the gradient of the gravitational force. Chappell et al.~\cite{Chappell_2013} derived the expression for the gravitational potential around a prism of uniform density. Poisson's equation for the gravitational field is:
\begin{equation}
-{\nabla}\cdot{}(\nabla\phi_G)=4\pi{}G\rho~,
\end{equation}
i.e. of the same type as Eq. \eqref{Eq.Poisson01}, but unlike this equation with no gradient on the right side of the equation. 

The expression found by Chappell et al.~is quite similar to the expression determined for the magnetic prism above, in the sense that it is cyclic permutable and contains only $\textrm{arctan}$ and natural logarithmic functions. Additionally, the gravitational potential for a polyhedron is given by Ref. \cite{Schuhmacher_2024}.  We note that the expression for the gravitational potential can be split into a vector depending only on the geometry of the object, which could be termed the `degravitation' vector, multiplied with the density.  This has already been explored in the case of computing the field in Fourier space \cite{Boulanger_2024}, where the analogy between magnetism and gravity was cleanly explored, as well as in Ref. \cite{Byzov_2022}.

We note here that additionally, since the gravitational field, $\bf{g}$, can be defined from the potential as $\bf{g}=-\nabla\phi_G$, one can derive a `degravitation' tensor. In Newtonian gravity, a field is generated by any object with a mass, $M$. The gravitational field $\mbf{g}$ at point $\mbf{r}$ from an extended body located at position $\mbf{r'}$ with a mass distribution, $\rho$, is given by: 

\begin{equation*}
\mathbf g= \int G  \rho(\mathbf r') \frac{\mathbf r- \mathbf r'}{|\mathbf r- \mathbf r'|^3} d V'~,
\end{equation*}
where the volume integral of the density, $\rho$, is equal to the mass of the object.

If the density of the object is uniform, we can write:
\begin{eqnarray*}
\mathbf g&=&  G \rho \int  \frac{\mathbf r- \mathbf r'}{|\mathbf r- \mathbf r'|^3} d V' \\
&=&G \rho~\mathbb{K}(\mathbf r, \mathbf r')~,
\end{eqnarray*}
where the field then only depends on the geometrical function, $\mathbb{K}$, which is the degravitation tensor and the density of the object, similarly to the calculation of the magnetic field. 

\section{Conclusion}
Through integration, we analytically solve Poisson's equation for the magnetic scalar potential generated by a uniformly magnetized rectangular prism to determine a closed-form solution. We show that the solution, which is given in terms of arctan and natural logarithmic functions, can be written as a demagnetization vector, containing all the geometric information, multiplied with the magnetization. We validate the analytical expression by comparing with a finite element simulation and show that these agree perfectly. Finally, we show that the concept of demagnetization tensors can be extended to gravitational objects, which also obey Poisson's equation.

\section*{Data statement}
The derived analytical expressions for the magnetic scalar potential for a prism are available through the MagTense framework \cite{MagTense}, available at \href{https://www.magtense.org}{www.magtense.org}. Furthermore, the data is also available at the data repository at Ref. \cite{Data_2025}.

\section*{Acknowledgment}
This research was supported by Villum Foundation Synergy project number 50091 entitled "Physics-aware machine learning" and by the Independent Research Fund Denmark, grant “Magnetic Enhancements through Nanoscale Orientation (METEOR)”, 1032-00251B.

\bibliographystyle{IEEEtran}

\begin{thebibliography}{10}
\providecommand{\url}[1]{#1}
\csname url@samestyle\endcsname
\providecommand{\newblock}{\relax}
\providecommand{\bibinfo}[2]{#2}
\providecommand{\BIBentrySTDinterwordspacing}{\spaceskip=0pt\relax}
\providecommand{\BIBentryALTinterwordstretchfactor}{4}
\providecommand{\BIBentryALTinterwordspacing}{\spaceskip=\fontdimen2\font plus
\BIBentryALTinterwordstretchfactor\fontdimen3\font minus \fontdimen4\font\relax}
\providecommand{\BIBforeignlanguage}[2]{{%
\expandafter\ifx\csname l@#1\endcsname\relax
\typeout{** WARNING: IEEEtran.bst: No hyphenation pattern has been}%
\typeout{** loaded for the language `#1'. Using the pattern for}%
\typeout{** the default language instead.}%
\else
\language=\csname l@#1\endcsname
\fi
#2}}
\providecommand{\BIBdecl}{\relax}
\BIBdecl

\bibitem{exl2020micromagnetism}
L.~Exl, D.~Suess, and T.~Schrefl, ``Micromagnetism,'' \emph{Handbook of Magnetism and Magnetic Materials}, pp. 1--44, 2020.

\bibitem{bjork_explaining_2023}
\BIBentryALTinterwordspacing
R.~Bj\o{}rk and A.~R. Insinga, ``\BIBforeignlanguage{en}{Explaining {Browns} paradox in {NdFeB} magnets from micromagnetic simulations},'' \emph{\BIBforeignlanguage{en}{Journal of Magnetism and Magnetic Materials}}, vol. 571, p. 170510, Apr. 2023. [Online]. Available: \url{https://www.sciencedirect.com/science/article/pii/S0304885323001592}
\BIBentrySTDinterwordspacing

\bibitem{mackay2006divergence}
F.~Mackay, R.~Marchand, and K.~Kabin, ``Divergence-free magnetic field interpolation and charged particle trajectory integration,'' \emph{Journal of Geophysical Research: Space Physics}, vol. 111, no.~A6, 2006.

\bibitem{bernauer2016measurement}
J.~Bernauer, J.~Diefenbach, G.~Elbakian, G.~Gavrilov, N.~Goerrissen, D.~Hasell, B.~Henderson, Y.~Holler, G.~Karyan, J.~Ludwig \emph{et~al.}, ``Measurement and tricubic interpolation of the magnetic field for the olympus experiment,'' \emph{Nuclear Instruments and Methods in Physics Research Section A: Accelerators, Spectrometers, Detectors and Associated Equipment}, vol. 823, pp. 9--14, 2016.

\bibitem{le20123}
E.~Le~Grand and S.~Thrun, ``3-axis magnetic field mapping and fusion for indoor localization,'' in \emph{2012 IEEE International Conference on Multisensor Fusion and Integration for Intelligent Systems (MFI)}.\hskip 1em plus 0.5em minus 0.4em\relax IEEE, 2012, pp. 358--364.

\bibitem{solin2018modeling}
A.~Solin, M.~Kok, N.~Wahlstr{\"o}m, T.~B. Sch{\"o}n, and S.~S{\"a}rkk{\"a}, ``Modeling and interpolation of the ambient magnetic field by gaussian processes,'' \emph{IEEE Transactions on robotics}, vol.~34, no.~4, pp. 1112--1127, 2018.

\bibitem{moresi2003miniature}
G.~Moresi and R.~Magin, ``Miniature permanent magnet for table-top nmr,'' \emph{Concepts in Magnetic Resonance Part B: Magnetic Resonance Engineering: An Educational Journal}, vol.~19, no.~1, pp. 35--43, 2003.

\bibitem{raich2004design}
H.~Raich and P.~Bl{\"u}mler, ``Design and construction of a dipolar halbach array with a homogeneous field from identical bar magnets: Nmr mandhalas,'' \emph{Concepts in Magnetic Resonance Part B: Magnetic Resonance Engineering: An Educational Journal}, vol.~23, no.~1, pp. 16--25, 2004.

\bibitem{pollok_magnetic_2023}
\BIBentryALTinterwordspacing
S.~Pollok, N.~Olden-J\o{}rgensen, P.~S. J\o{}rgensen \emph{et~al.}, ``Magnetic {Field} {Prediction} {Using} {Generative} {Adversarial} {Networks},'' \emph{Journal of Magnetism and Magnetic Materials}, vol. 571, p. 170556, Apr. 2023. [Online]. Available: \url{http://arxiv.org/abs/2203.07897}
\BIBentrySTDinterwordspacing

\bibitem{Bischof_2025}
R.~Bischof and M.~A. Kraus, ``Multi-objective loss balancing for physics-informed deep learning,'' \emph{Computer Methods in Applied Mechanics and Engineering}, vol. 439, p. 117914, 2025.

\bibitem{Rhodes_1954}
P.~Rhodes, ``Demagnetising energies of uniformly magnetised rectangular blocks,'' \emph{Proc. Leeds Phil. Liter. Soc}, vol.~6, pp. 191--210, 1954.

\bibitem{Hubert_book}
A.~Hubert and R.~Sch{\"a}fer, \emph{Magnetic domains: the analysis of magnetic microstructures}.\hskip 1em plus 0.5em minus 0.4em\relax Springer Science \& Business Media, 2008.

\bibitem{Schabes_2003}
M.~Schabes and A.~Aharoni, ``Magnetostatic interaction fields for a three-dimensional array of ferromagnetic cubes,'' \emph{IEEE Transactions on Magnetics}, vol.~23, no.~6, pp. 3882--3888, 2003.

\bibitem{joseph_1964}
\BIBentryALTinterwordspacing
R.~I. Joseph and E.~Schl\"{o}mann, ``Demagnetizing {Field} in {Nonellipsoidal} {Bodies},'' \emph{Journal of Applied Physics}, vol.~36, no.~5, pp. 1579--1593, Nov. 1964. [Online]. Available: \url{https://doi.org/10.1063/1.1703091}
\BIBentrySTDinterwordspacing

\bibitem{Smith_2010}
A.~Smith, K.~K. Nielsen, D.~Christensen, C.~R.~H. Bahl, R.~Bj{\o}rk, and J.~Hattel, ``The demagnetizing field of a nonuniform rectangular prism,'' \emph{Journal of Applied Physics}, vol. 107, no.~10, p. 103910, 2010.

\bibitem{Nielsen_2020}
K.~K. Nielsen and R.~Bj{\o}rk, ``The magnetic field from a homogeneously magnetized cylindrical tile,'' \emph{Journal of Magnetism and Magnetic Materials}, vol. 507, p. 166799, 2020.

\bibitem{Slanovc_2022}
F.~Slanovc, M.~Ortner, M.~Moridi, C.~Abert, and D.~Suess, ``Full analytical solution for the magnetic field of uniformly magnetized cylinder tiles,'' \emph{Journal of Magnetism and Magnetic Materials}, vol. 559, p. 169482, 2022.

\bibitem{Joseph_1967}
R.~Joseph, ``Ballistic demagnetizing factor in uniformly magnetized rectangular prisms,'' \emph{Journal of applied physics}, vol.~38, no.~5, pp. 2405--2406, 1967.

\bibitem{Aharoni_1998}
A.~Aharoni, ``Demagnetizing factors for rectangular ferromagnetic prisms,'' \emph{Journal of Applied Physics}, vol.~83, no.~6, pp. 3432--3434, 1998.

\bibitem{Fukushima_1998}
H.~Fukushima, Y.~Nakatani, and N.~Hayashi, ``Volume average demagnetizing tensor of rectangular prisms,'' \emph{IEEE Transactions on Magnetics}, vol.~34, no.~1, pp. 193--198, 1998.

\bibitem{Jackson}
J.~D. Jackson, \emph{Classical Electrodynamics, 3rd ed.}\hskip 1em plus 0.5em minus 0.4em\relax Hoboken, NJ, USA: Wiley, 1999.

\bibitem{MagTense}
R.~Bjørk and K.~K. Nielsen, ``Magtense - a micromagnetism and magnetostatic framework,'' \emph{doi.org/10.11581/DTU:00000071, https://www.magtense.org}, 2019.

\bibitem{Boulanger_2024}
O.~Boulanger, ``2d fast fourier transform analytical solutions in all space for all gravity and magnetic components,'' \emph{Geophysical Prospecting}, vol.~72, no.~2, pp. 809--832, 2024.

\bibitem{Beleggia_2005}
M.~Beleggia, M.~De~Graef, Y.~T. Millev, D.~A. Goode, and G.~Rowlands, ``Demagnetization factors for elliptic cylinders,'' \emph{Journal of Physics D: Applied Physics}, vol.~38, no.~18, p. 3333, 2005.

\bibitem{Nielsen_2019}
K.~K. Nielsen, A.~R. Insinga, and R.~Bj{\o}rk, ``The stray and demagnetizing field of a homogeneously magnetized tetrahedron,'' \emph{IEEE Magnetics Letters}, vol.~10, pp. 1--5, 2019.

\bibitem{osborn_1945}
J.~A. Osborn, ``Demagnetizing factors of the general ellipsoid,'' \emph{Physical review}, vol.~67, no. 11-12, p. 351, 1945.

\bibitem{Tejedor_1995}
M.~Tejedor, H.~Rubio, L.~Elbaile, and R.~Iglesias, ``External fields created by uniformly magnetized ellipsoids and spheroids,'' \emph{IEEE transactions on magnetics}, vol.~31, no.~1, pp. 830--836, 1995.

\bibitem{Beleggia_2009}
M.~Beleggia, D.~Vokoun, and M.~De~Graef, ``Demagnetization factors for cylindrical shells and related shapes,'' \emph{Journal of Magnetism and Magnetic Materials}, vol. 321, no.~9, pp. 1306--1315, 2009.

\bibitem{PratCamps_2016}
J.~Prat-Camps, C.~Navau, A.~Sanchez, and D.-X. Chen, ``Demagnetizing factors for a hollow sphere,'' \emph{IEEE Magnetics Letters}, vol.~7, pp. 1--4, 2015.

\bibitem{Chen_2002}
D.-X. Chen, E.~Pardo, and A.~Sanchez, ``Demagnetizing factors of rectangular prisms and ellipsoids,'' \emph{IEEE Transactions on Magnetics}, vol.~38, no.~4, pp. 1742--1752, 2002.

\bibitem{Chen_2005}
------, ``Demagnetizing factors for rectangular prisms,'' \emph{IEEE Transactions on magnetics}, vol.~41, no.~6, pp. 2077--2088, 2005.

\bibitem{comsol}
\BIBentryALTinterwordspacing
S.~COMSOL~AB, Stockholm. (2023) Comsol multiphysics® v. 6.2. [Online]. Available: \url{www.comsol.com}
\BIBentrySTDinterwordspacing

\bibitem{Chappell_2013}
J.~M. Chappell, M.~J. Chappell, A.~Iqbal, and D.~Abbott, ``The gravity field of a cube,'' \emph{Physics International}, vol.~3, no.~2, pp. 50--57, 2013.

\bibitem{Schuhmacher_2024}
J.~Schuhmacher, E.~Blazquez, F.~Gratl, D.~Izzo, and P.~G{\'o}mez, ``Efficient polyhedral gravity modeling in modern c++ and python,'' \emph{Journal of Open Source Software}, vol.~9, no.~98, p. 6384, 2024.

\bibitem{Byzov_2022}
D.~Byzov, P.~Martyshko, and A.~Chernoskutov, ``Computationally effective modeling of self-demagnetization and magnetic field for bodies of arbitrary shape using polyhedron discretization,'' \emph{Mathematics}, vol.~10, no.~10, p. 1656, 2022.

\bibitem{Data_2025}
\BIBentryALTinterwordspacing
R.~Bj\o{}rk, ``Data for the article ``the magnetic scalar potential for a rectangular prism'','' \emph{data.dtu.dk, DOI:10.11583/DTU.27879888}, 2024. [Online]. Available: \url{https://doi.org/10.11583/DTU.27879888}
\BIBentrySTDinterwordspacing

\end{thebibliography}

\end{document}